\documentclass{article}
\usepackage{graphicx}
\usepackage{url}
\usepackage[authoryear]{natbib}

\providecommand{\keywords}[1]
{
  \small	
  \textbf{\textit{Keywords---}} #1
}

\title{Recency predicts bursts \\ in the evolution of author citations}

\author{Filipi Nascimento Silva$^{1}$, Aditya Tandon$^2$, \\ Diego Raphael Amancio$^3$,  Alessandro Flammini$^{1,2}$, Filippo Menczer$^{1,2}$,\\ Sta\v{s}a Milojevi\'{c}$^2$ \& Santo Fortunato$^{1,2}$}

\date{}

\begin{document}

\maketitle
{\small
\centering
$^{1}$ Indiana University Network Science Institute, USA  

$^2$ Center for Complex Networks and Systems Research,\\Luddy School of Informatics, Computing and Engineering,\\Indiana University, Bloomington, USA

$^3$ Institute of Mathematics and Computer Science,\\University of S\~{a}o Paulo,
S\~{a}o Carlos, Brazil

}

\begin{abstract}
The citations process for scientific papers has been studied extensively. But while the citations accrued by authors are the sum of the citations of their papers, translating the dynamics of citation accumulation from the paper to the author level is not trivial. Here we conduct a systematic study of the evolution of author citations, and in particular their bursty dynamics. We find empirical evidence of a correlation between the number of citations most recently accrued by an author and the number of citations they receive in the future. Using a simple model where the probability for an author to receive new citations depends only on the number of citations collected in the previous 12--24 months, we are able to reproduce both the citation and burst size distributions of authors across multiple decades.
\end{abstract}

\vskip1cm

\section{Introduction}

Citations are one of the most widely used indicators of academic impact and, as such, they have been studied extensively~\citep{WALTMAN2016365}. 
Despite a lack of consensus about the relevance of citations as an indicator of quality~\citep{leydesdorff2016,martin1983},
papers and authors with a large number of citations are considered influential. Understanding the process of citation accumulation is one of the central questions in science of science~\citep{fortunato18}. The major challenge lies in delineating how the interplay between factors related to the quality and relevance of papers and factors related to author popularity contribute to the process of citation accumulation. 

The first model of citation dynamics for papers was proposed by \cite{price76}. It is based on the principle of \textit{cumulative advantage}: the probability of a paper to be cited is proportional to the number of citations the paper already has, up to an additive constant. This principle leads to a broad distribution of citations: most papers have just a few citations, while a minority of top-cited papers accounts for a considerable fraction of all citations~\citep{price65,radicchi08,thelwall16}.

In network science~\citep{newman10,barabasi16} the principle of cumulative advantage is called \textit{preferential attachment} and it has been invoked to explain the broad degree distributions observed in many real networks~\citep{barabasi99}.
The phenomenon is also known as \textit{rich-get-richer} or \textit{Matthew effect} in the sociology of science, where certain psycho-social processes lead the community to give disproportionately large credit to individuals that already enjoy a high reputation~\citep{merton68}.  These dynamics have been argued to lead to inequalities or stratification in science~\citep{cole1974,zuckerman1977,diprete2006} and the existence of star scientists~\citep{moody2004}, though the process itself is not straightforward~\citep{allison1982}.

In the simplest models of paper citation dynamics based on preferential attachment, every paper keeps accumulating citations forever, although at a slowing rate due to the increasing competition with newly published papers. It is well known, however, that most papers have a finite lifetime, so that most citations are accrued within the first few years after publication and the probability of being cited often dramatically decreases thereafter~\citep{stringer08, parolo15, hajra05,eom11,wang13} --- with some notable exceptions~\citep{ke2015}.
This reflects the obsolescence of knowledge, in that attention shifts from old findings to newer ones, which become the basis of future research. A related consequence is the \textit{recency effect}, i.e., the fact that the probability of receiving new citations is somewhat dependent on the citations collected in recent times~\citep{golosovsky12,wang2008}.

By including obsolescence and recency, as well as other ingredients, models can successfully describe the citation dynamics of papers~\citep{eom11,golosovsky12}, to the point that it is possible to predict the future citation trajectory of individual papers~\citep{wang13}.  

Despite all these advances in our understanding of \emph{paper} citation dynamics, \emph{author} citation dynamics have received little attention in the literature.
On the empirical side, this is mostly due to the challenges related to author name disambiguation~\citep{Ferreira:2012:BSA:2350036.2350040}. On the theoretical side, in principle, our understanding of citation accumulation for papers could be leveraged to characterize and model the citation dynamics of authors: the citation count of an author, after all, is the sum of the citation counts of their papers. Nevertheless, models based on publication portfolios would involve many parameters and assumptions, including paper lifetimes, author productivity, and how productivity is related to author success and number of citations. 

In this paper, we characterize and model the process of citation accumulation for authors.  We focus on two quantitative signatures: the distributions of the number of citations and of the size of \textit{citation bursts}. As it happens for papers~\citep{eom11}, both distributions are broad. The fact that the burst size distribution is heavy-tailed is incompatible with a dynamics driven by preferential attachment alone. We find that both distributions can be well described by a simple model whose sole driver is the number of recent citations.

\section{Results}

Our analysis is based on a data set of 577$,$870 papers published in 15 journals of the American Physical Society (APS, \url{journals.aps.org/datasets}), from 1893 until 2015 (see Table S1 in Supplementary Information). 

When considering the list of authors of each paper in the dataset, a major hurdle is that author names can be ambiguous --- multiple authors can have the same name and multiple names can be used by the same author. The recently created Microsoft Academic Graph (MAG) is a large publications database encompassing all scientific disciplines, which uses sophisticated machine learning algorithms to disambiguate author names~\citep{MAG}. By mapping APS papers onto the MAG we were able to assign all papers to a set of by 732$,$965 disambiguated authors.

\subsection{Author Citations} 

\begin{figure}[t]
\centering
\includegraphics[width=\linewidth]{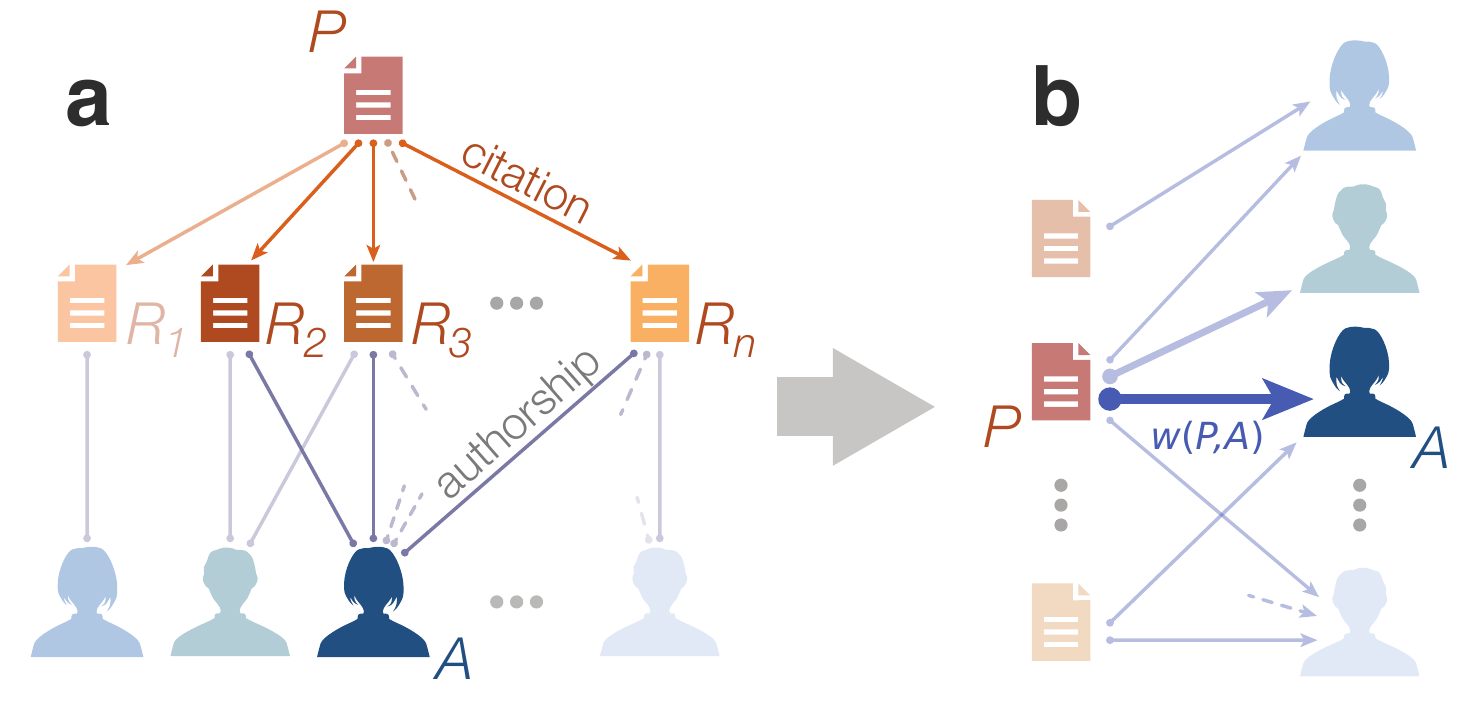}
\caption{Bipartite Paper-Author citation network. (a) A paper $P$ in our data set cites articles ($R_1$, $R_2$, $\dots$, $R_{N}$). The orange lines represent citations between papers, the blue lines match each author to their papers. (b) From the paper-paper citations we derive the citations between papers and authors, yielding a weighted bipartite network. 
\label{fig:schematic}}
\end{figure}

We use the APS data set to build a bipartite paper-author citation network (BPAN). For each citation from a paper $P$ to a paper $R_i$, we set a direct link going from $P$ to each author $A$ of $R_i$. The weight of each link $w(P,A)$ corresponds to the number of articles coauthored by $A$ that are cited by $P$.  The number of citations of author $A$ is the sum of $w(P,A)$ over all papers $P$ citing $A$. 
Fig.~\ref{fig:schematic} illustrates the process of generating a BPAN from the paper citation network.

We studied the evolution of the number of citations received by authors over a long time span, between 1930 and 2010. When we refer to a specific year $t$ we mean the set of all authors publishing papers from the beginning of the APS history (1893) along with all their mutual citations until year $t$. 

\begin{figure}[t]
\centering
\includegraphics[width=0.85\linewidth]{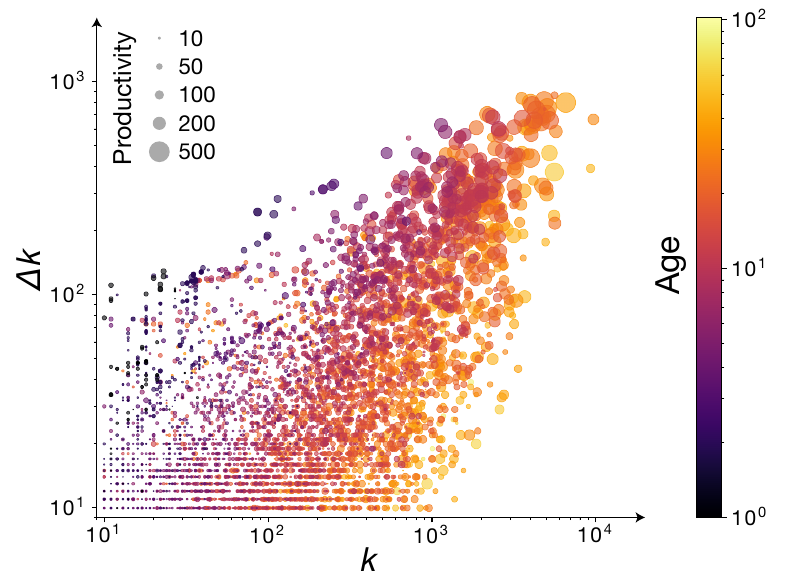}
\caption{Relationship between the total number of citations received by an author $i$ up until 2010, $k=k_i^{2009}$, and the citations received by the same author in 2010, $\Delta k = k_i^{2010}-k_i^{2009}$. The academic age of authors is represented by color and their productivity up to 2010 as symbol size. For clarity purposes, the plot was constructed from a random sample of $10\%$ of the authors in the data set, and focuses on authors with $k \ge 10$ and $\Delta k \ge 10$.
\label{fig:kvsDk}}
\end{figure}

In Fig.~\ref{fig:kvsDk} we show the relation between the number of citations $\Delta k$ received by an author in 2010 and the number of citations $k$ received in all previous years. The diagram shows that author citation dynamics are \textit{bursty}: the increment $\Delta k$ can vary by orders of magnitude among authors having the same total number of citations. We observe a clear correlation between $k$ and $\Delta k$ but also a large dispersion. Large values of $\Delta k$ tend to be associated to authors with higher career age and productivity, but they are not unusual among early-career scholars. Such a bursty character of author citation dynamics is the main focus of this paper.

\begin{figure}
\centerline{\includegraphics[width=0.9\textwidth]{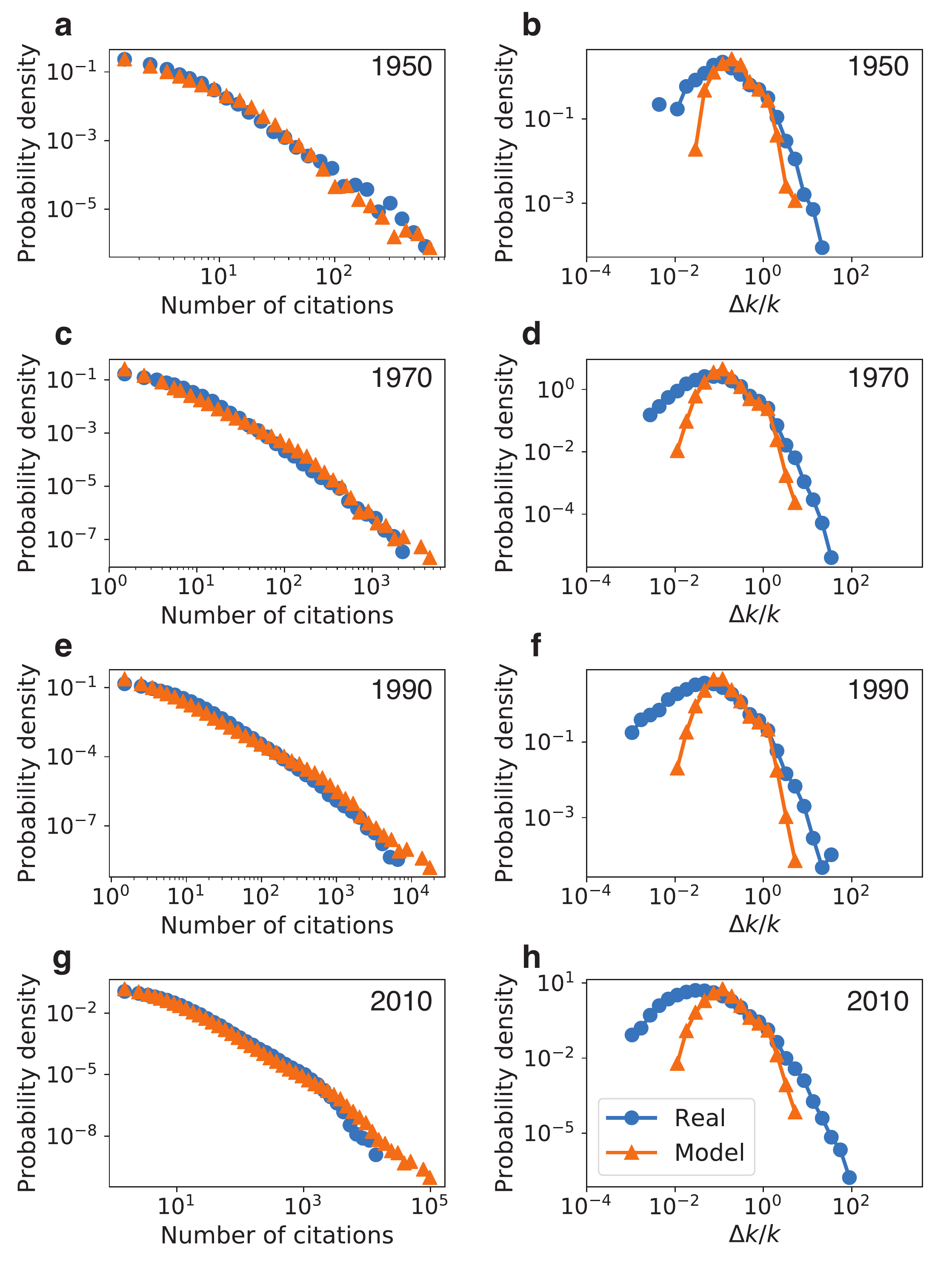}}
\caption{Empirical distributions in BACNs (circles). Citation distributions (left) are computed for all authors and mutual citations from the beginning of the dataset (1893) with the model starting in 1930 and simulated until (a)~1950, (c)~1970, (e)~1990, and (g)~2010. Burst size distributions (right) are computed by considering the increments $\Delta k$ in the number of citations of all authors in the year (b)~1950, (d)~1970, (f)~1990, and (h)~2010. Both distributions span multiple orders of magnitude. A simple model based on pure preferential attachment (triangles) is able to reproduce the heavy-tailed citation distributions, while it generates much narrower burst size distributions, indicating that the predicted increments do not have high variability.}
\label{empdistr}
\end{figure}

\begin{figure}
\centerline{\includegraphics[width=\linewidth]{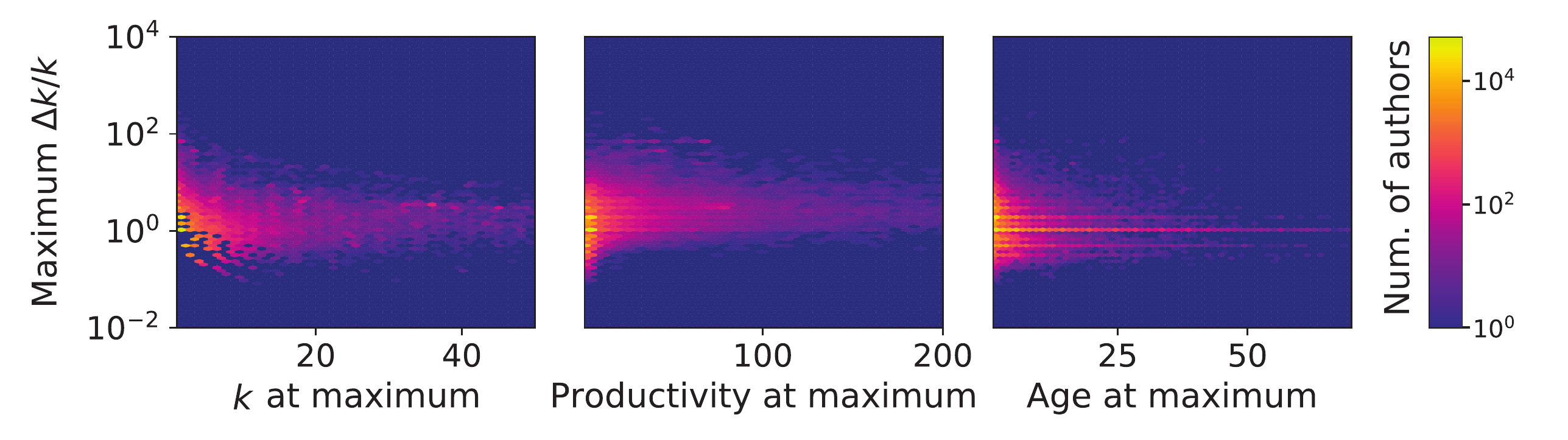}}
\caption{Distribution of maximum burstiness among authors according to their age, number of citations, and productivity at the peak.}
\label{figmaxburst}
\end{figure}

Let us consider the distributions of two variables. The first variable is the \textit{number of citations} of an author. In Fig.~\ref{empdistr}(a,c,e,g) we see that the distribution is broad, as expected: most authors are poorly cited, whereas a few receive many citations. The second variable is the \textit{citation burst size}, which is computed as follows. Given some reference year $t$, for each author $i$ we compute the number of their citations until years $t-1$ and $t$, which we indicate as $k_i^{t-1}$ and $k_i^{t}$. The burst size at year $t$ is then defined as the ratio between the number of citations collected in year $t$ and the number of citations until the year before:
\begin{equation} 
\label{eq:burst}
b_i^{t}=\frac{\Delta k_i^t}{k_i^{t-1}}=\frac{k_i^{t}-k_i^{t-1}}{k_i^{t-1}}.
\end{equation}
The distribution of citation burst sizes is shown in Fig.~\ref{empdistr}(b,d,f,h). This distribution is broad as well, as already observed in paper citation dynamics~\citep{eom11}. With very low probability, authors may receive in a single year up to 100 times the number of citations they received in their entire career up to the beginning of that year. This is the same trend observed at the paper level~\citep{eom11} and also in the dynamics of popularity~\citep{ratkiewicz10}. 
While the largest bursts occur more often in the initial phase of a scholar's career, when the number of papers and the corresponding citation counts are relatively low, large bursts can also occur at later times (Fig.~\ref{figmaxburst}).

Abrupt increments in the number of citations might signal a sudden increase in the productivity of the author, the beginning of a ``hot streak" with the publication of papers of significantly higher impact than earlier output~\citep{liu18}, or a ``sleeping beauty" paper that starts receiving a lot of credit from the author's peers~\citep{ke2015}. 
The shapes of the burst size distributions are robust across the years and, as such, deserve a general explanation.

\subsection{Model Implementation}

Our model of author evolution starts in a reference year $t_{in}$ and consider the following one-month periods until a final year $t_{f}$. For each month, we add new papers published in that month and their authors, together with their citations to existing authors. We track the number of citations $k_j^t$ received by each author $j$ in each month $t$.

For each paper $p$ published in a given month $t$, we consider all authors of $p$. New authors are added to the system. The number of authors $c_p$ cited by $p$ includes multiple citations to the same author that originate from distinct references. We add $c_p$ citations from $p$ to existing authors according to some rule specific to the particular model. 

At each stage of the evolution, the model system has the same number of authors and total number of citations as the actual system. We measure empirical distributions of citations $k$ and burstiness $\Delta k/k$ for each year. We would like to explain the shapes of the empirical distribution by reproducing them via simple citation rules.

\subsection{Preferential Attachment} 

First, we consider a simple preferential attachment rule. The probability that author $j$ receives a citation in an interval of time starting at $t$ depends linearly on the number of citations $k_j$ they have received until that time:
\begin{equation} 
\label{eq:PA}
P(i \rightarrow j) \propto A+k_j^t.
\end{equation}
The constant $A>0$ attributes a non-zero probability to receive citations to authors that have received none so far. Eq.~\ref{eq:PA} defines Price's model of citation dynamics~\citep{price76}.
In Fig.~\ref{empdistr} we compare the empirical distributions with those produced by this model  (see Methods). The model uses $A=1.8$, a value that was chosen by fitting the distribution of the number of citations. 
The model reproduces the profiles of the citation distributions, which exhibit progressively broader support the longer the simulation runs. For 2010 the model curve stretches one order of magnitude further than the empirical curve. This is because the model ignores any factor related to obsolescence: authors never stop receiving citations according to preferential attachment and their total can become arbitrarily large if one waits sufficiently long.  

The burst size distribution generated by the model is much narrower than the empirical one. According to preferential attachment (Eq.~\ref{eq:PA}), the increment in the number of citations of an author in a given (small) time window should be approximately proportional to the number of citations collected before, so the ratio $\Delta k/k$ should be roughly constant. In fact, the bell-shaped model distribution for the burst size represents random Poissonian fluctuations about the mean. The discrepancy between model and data becomes more pronounced the longer the dynamics run. It is thus apparent that preferential attachment alone cannot account for the bursty citation dynamics we observe for authors, as already seen for papers~\citep{eom11}.

\subsection{Recency} 

The success of an author is the success of their papers. 
Papers have a finite lifetime~\citep{stringer08, parolo15, hajra05,eom11,wang13} and collect a significant fraction of all their citations in a limited interval of time, although rare exceptions of \textit{evergreen} papers exist~\citep{ZHANG2017629}.

\begin{figure}[t]
\centering
\includegraphics[width=0.80\linewidth]{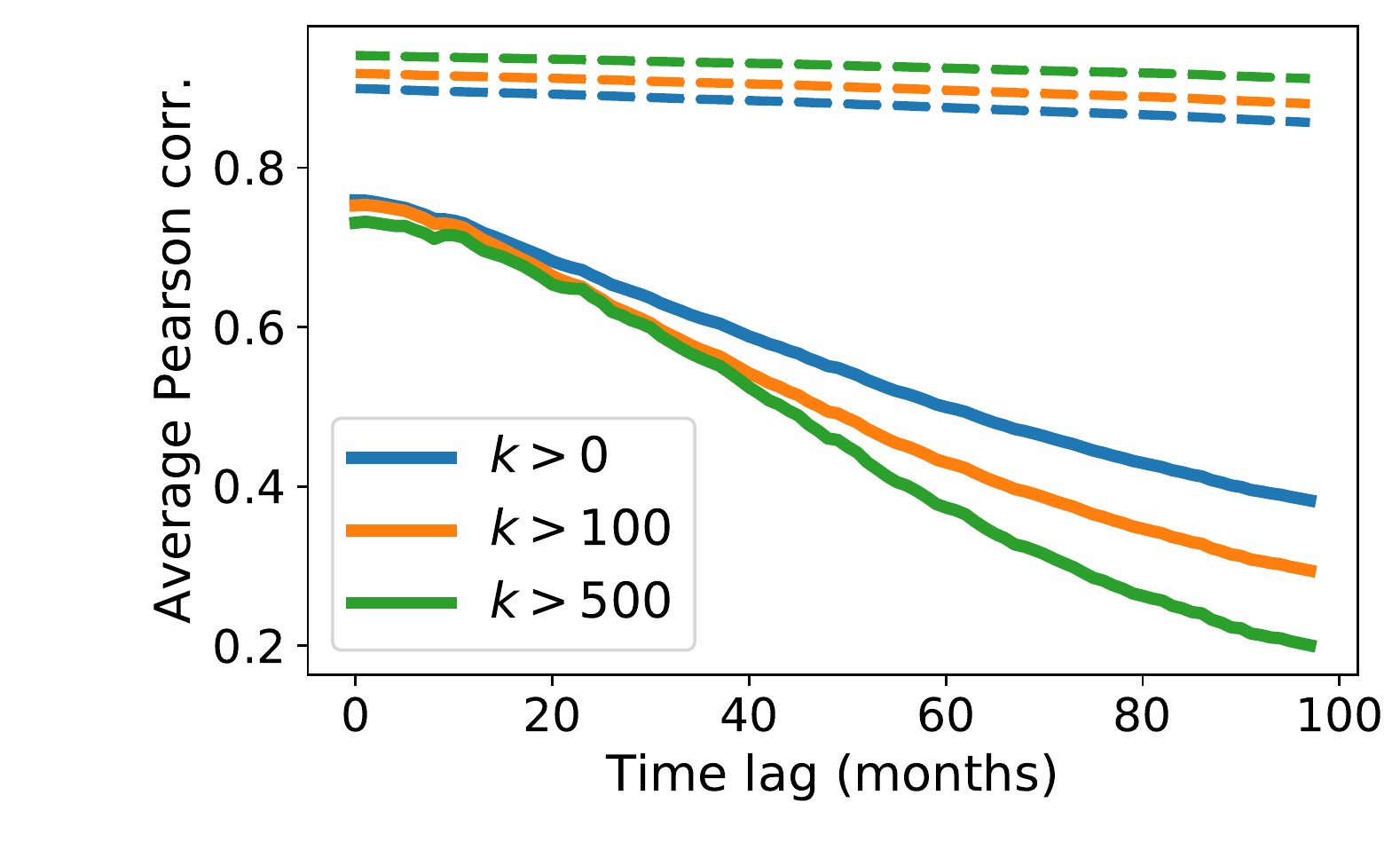}
\caption{Recency in author citation dynamics. We show the Pearson correlation coefficient between the number of citations accrued by an author in a given month $t$ and the number of citations obtained in month $t-w$, with $w=1, 2, 3, \dots, 100$. 
The blue line is the result when all authors are considered, regardless of their number of citations. The orange and the green lines correspond to authors having more than $100$ and $500$ citations, respectively, at time $t$. Curves are averaged over $t$, with $t$ being each month in the 10-year period (2000--2010). The dashed lines show the correlation obtained by the simple preferential attachment model, which decreases very slowly with lag.
}
\label{figcorrt}
\end{figure}

In most cases, the number of citations collected by a paper in a given interval varies smoothly over time, so there is a sizeable correlation between the number of citations in nearby intervals~\citep{golosovsky12,wang2008}. 
 Such recency effect occurs for authors as well.
 One can only speculate on the reasons behind this. For example, according to~\cite{liu18}, this could be due to \textit{hot streaks}, the ability to produce papers with sustained impact within short periods of time. 
It is therefore plausible to assume recency because of the inertia in the citation increments of individual papers and because impactful papers are likely to appear in sequence. 
In Fig.~\ref{figcorrt} we show the correlation between the numbers of monthly citations received by an author $w$ months apart. We see that the correlation is important and slowly decreases with $w$. For highly cited authors the correlation decreases faster. We conclude that recency 
plays an important role in author citation dynamics.

\subsection{Recency Model} 

We test a rule originally introduced by \cite{wang2008}, which, although inspired by preferential attachment, gives more weight to citations received recently in the determination of the probability to receive new citations in the  future. The probability that author $j$ receives a new citation at time $t$ is proportional to
\begin{equation}
   P(i \rightarrow j) \propto A+\Delta k_j^{[t, t-w]},
\end{equation}
where $A$ is an additive constant and $\Delta k_j^{[t, t-w]}= k_j^t - k_j^{t-w} $ is the number of citations that $j$ has accrued in the previous $w$ months. The model has thus two parameters: $A$ and $w$.

\begin{figure}
\centerline{\includegraphics[width=\linewidth]{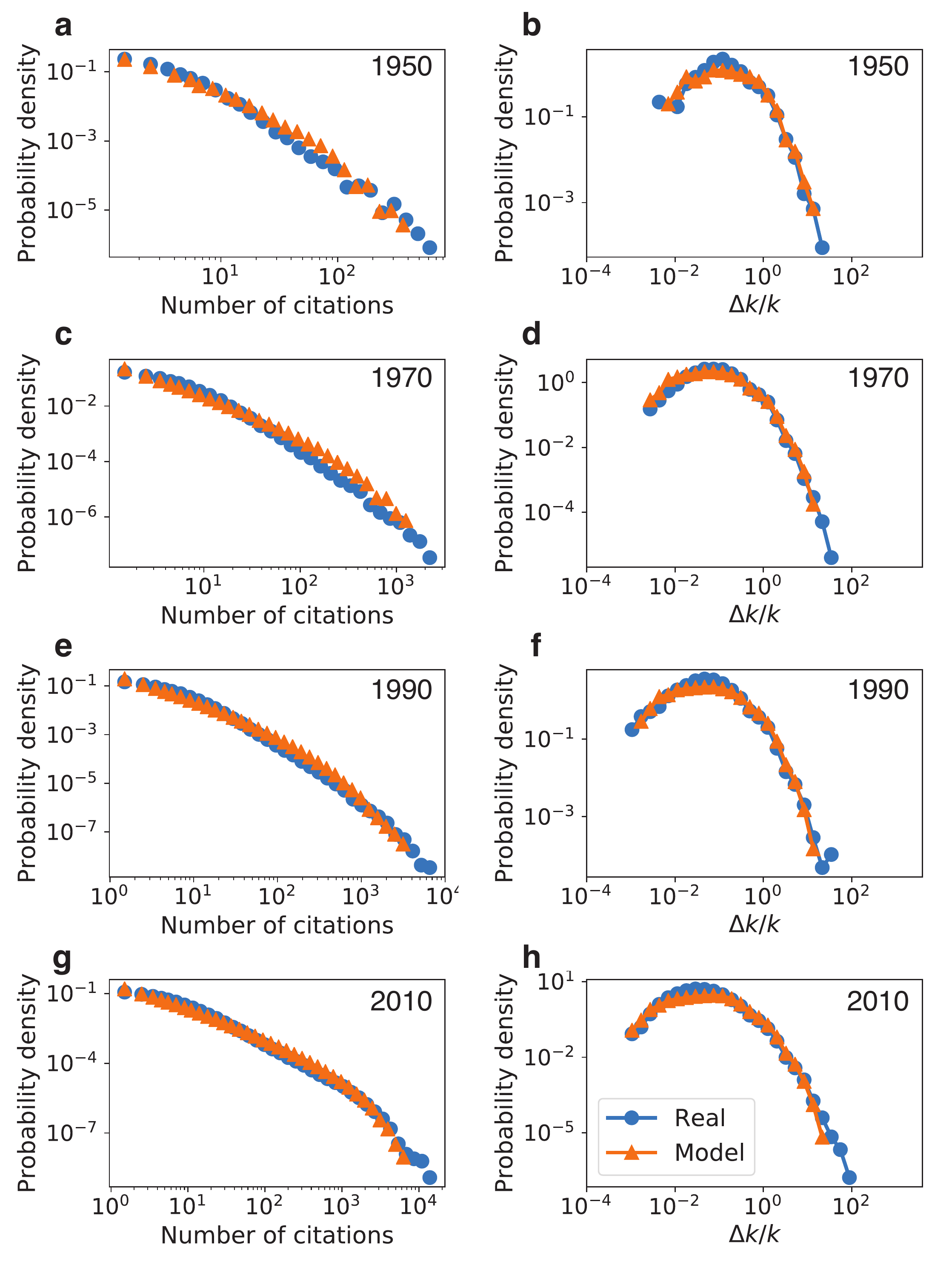}}
\caption{Comparison between the recency model and the data. The empirical distributions are the same as in Fig.~\ref{empdistr}. The model closely follows both empirical curves throughout the evolution.}
\label{figrec}
\end{figure}

Fig.~\ref{figrec} compares the empirical distributions of Fig.~\ref{empdistr} with those obtained from the recency model, with best-fit values for the parameters $A$ and $w$. We see that the recency model describes both distributions well throughout the period (1950--2010). In Supplementary Information (Fig.~S1) we show the comparison between model and data when the dynamics start from the actual configuration of APS authors as of 1970, with all actual citations each author collected until then. 

\begin{figure}
\centerline{\includegraphics[width=\linewidth]{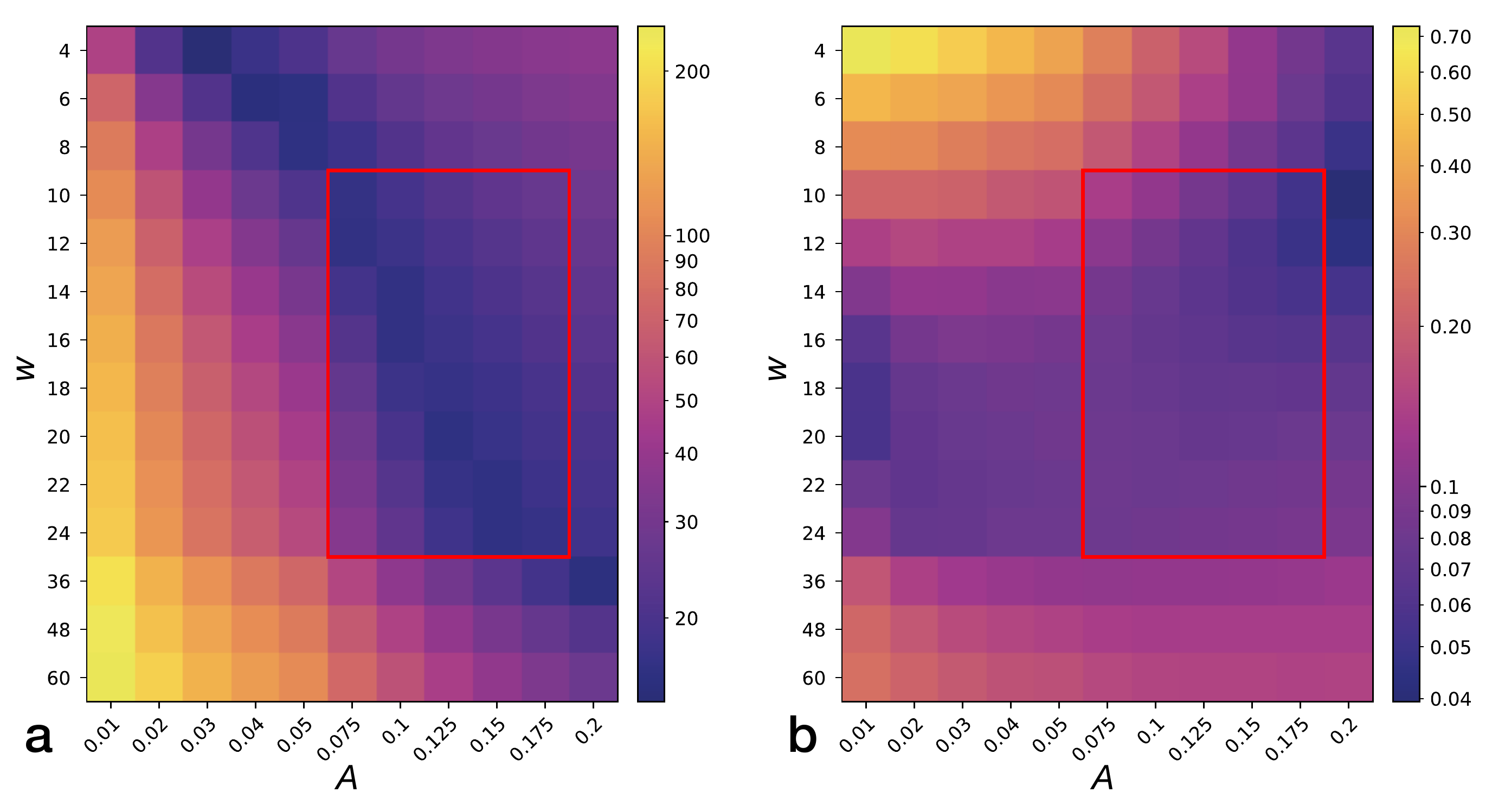}}
\caption{Map of the Wasserstein distance obtained between model and empirical distributions for (a)~citations and (b)~burst sizes considering many pairs of parameters $A$ and $w$. The region highlighted in red corresponds to the range of parameters resulting in the best compromise between the qualities of the fits for the citation and burst size distributions. Fig.~S2 in Supplementary Information shows the actual curves obtained for each parameter configuration.}
\label{figpar}
\end{figure}

In Fig.~\ref{figpar} we show the goodness of fit of both distributions for different parameter choices, using the Wasserstein distance. The parameter ranges leading to the best fits is highlighted.
Remarkably, all model curves shown in Fig.~\ref{figrec} correspond to the same pair of values of the parameters: $A=0.075$ and $w=12$. But values of $w$ ranging from 12 to 24 months lead to fits of comparable quality. 
Therefore, we conclude that the number of citations accrued by an author in the last one-two years is an important driver of the dynamics. In fact, this ingredient alone is capable of providing a good description of both  citation and burst size distributions for 80 years of APS author citation evolution. 

\section{Discussion}

We have studied the evolution of the citation dynamics of APS authors. As observed for papers, the citation distribution is broad and the dynamics are bursty, in that the number of citations collected by an author in a given interval can have sharp fluctuations. Also, we find a strong correlation between the numbers of citations accrued in nearby time intervals, confirming that recency is an important factor in the dynamics. Indeed, a model based on recency alone suffices to account for both the citation distribution and the burstiness of the dynamics, over eight decades of evolution. The best match between the model and empirical curves suggests that the key driver is the number of citations received by an author over the last 12--24 months. We could thus claim that an author is as ``hot'' as they have been in the last one-two years. 

We stress that our work focuses on the outcomes of the dynamics at the author population level. Moving to the more ambitious goal of describing and even predicting citation trajectories for individual authors remains an open challenge that will likely require the introduction of additional ingredients into the model~\citep{liu18}.

\paragraph{Acknowledgments.} We thank Xiaoran Yan for precious assistance with author disambiguation. This work uses publication data from the American Physical Society and Microsoft Academic Graph data by Microsoft Research provided by the Indiana University Network Science Institute. We gratefully acknowledge support from the US Navy (award N00174-17-1-0007), US AFOSR (Minerva awards FA9550-19-1-0391 and FA9550-19-1-0354), FAPESP (grants 2015/08003-4, 2017/09280-7 and 2017/13464-6), and CNPq-Brazil (grant 304026/2018-2).

\bibliographystyle{apalike}
\bibliography{comm}

\end{document}